# Weak Antilocalization Effect and Noncentrosymmetric Superconductivity in a Topologically Nontrivial Semimetal LuPdBi

Guizhou Xu[1], Wenhong Wang[1*], Xiaoming Zhang[1], Yin Du[1], Enke Liu[1], Shouguo Wang[1], Guangheng Wu[1], Zhongyuan Liu [2], and Xi Xiang Zhang[3]

[1] State Key Laboratory for Magnetism, Beijing National Laboratory for Condensed Matter Physics, Institute of Physics, Chinese Academy of Sciences, Beijing 100190, China

[2] State Key Laboratory of Metastable Material Sciences and Technology, Yanshan University, Qinhuangdao 066004, P. R. China

[3] Division of Physical Science and Engineering and Core Labs, King Abdullah University of Science and Technology (KAUST), Thuwal 23955-6900, Saudi Arabia

**A large number of half-Heusler compounds have been recently proposed as three-dimensional (3D) topological insulators (TIs) with tunable physical properties. However, no transport measurements associated with the topological surface states have been observed in these half-Heusler candidates due to the dominating contribution from bulk electrical conductance. Here we show that, by reducing the mobility of bulk carriers, a two-dimensional (2D) weak antilocalization (WAL) effect, one of the hallmarks of topological surface states, was experimentally revealed from the tilted magnetic field dependence of magnetoconductance in a topologically nontrivial semimetal LuPdBi. Besides the observation of a 2D WAL effect, a superconducting transition was revealed at $T_c \sim 1.7$ K in the same bulk LuPdBi. Quantitative analysis within the framework of a generalized BCS theory leads to the conclusion that the noncentrosymmetric superconductivity of LuPdBi is fully gapped with a possibly unconventional pairing character. The co-existence of superconductivity and the transport signature of topological surface states in the same bulk alloy suggests that LuPdBi represents a very promising candidate as a topological superconductor.**

PACS numbers: 74.70.Dd, 74.25.fc, 72.15.Rn, 73.23.-b

Keywords: Topological insulators, Heusler compounds, Weak antilocalization effect, Superconductivity

-------------------------------------------------------------------

[*] corresponding author. wenhong.wang@iphy.ac.cn

Topological insulators (TIs) with time-reversal-symmetry protected helical surface states have been identified as a new class of quantum materials for exploiting exciting physics such as Majorana fermions and quantum Hall effect, as well as developing potential applications in quantum computing.[1-5] Very recently, a number of ternary half-Heusler compounds of XYZ composition (X and Y are transition or rare-earth elements and Z is a heavy metal) have been predicted theoretically to be three-dimensional (3D) TI candidates due to their topologically nontrivial band inversion.[6-8] In particular, some of half-Heusler TI candidate compounds containing rare-earth elements with strongly correlated $f$ electrons already exhibit many extreme properties, such as magnetism,[9] heavy fermion behavior,[10] and superconductivity,[11-15] which give rise to the likelihood of searching topological superconductors[16-20] with a fully gapped and odd-parity Cooper pairing states in this class of materials. Despite its importance, however, making observations of the signature of the surface states through the electrical transport measurements in these half-Heusler TI candidate compounds is challenging as a result of the dominating contribution from bulk electrical conductance. Consequently, the reported large linear magnetoresistance and Shubnikov-de Hass quantum oscillations in half-Heusler TIs have been ascribed to the high mobility of bulk carriers.[21,22] Therefore, reducing the mobility of the charge carriers in the bulk is critical to revealing the signature of the surface states of TIs amongst the transport properties.[23-25] As the "clean" topological surface transport is expected to have high mobility with Dirac-cone like dispersion, when coexisted with the low mobility bulk conduction, it can reveal itself by quantum-featured transport properties. In this work, LuPdBi was chosen as a candidate for studying TIs and other properties particularly for the following reasons. On the one hand, based on the band structure calculations, as presented in **Figure 1,** we found that the highly dispersive $s$-like Bi $6s$ band locates well below the Fermi level $E_F$, whereas a less dispersive band consisting mainly of a Bi $6p$ band and a Lu $5d$ band is located just above the $E_F$ (compare ref. [26] and references therein). In such a case, the electrical transport of LuPdBi is dominated by the carriers from less dispersive $pd$ orbitals and, therefore, greatly degrades the mobility of the bulk carriers. On the other hand, the less dispersive band exhibits a Van Hove singularity along the Γ-X direction across the $E_F$. In fact, this Van Hove scenario was used to explain the superconductivity in Heusler compounds.[26-30] Here, in this paper, we report magneto-transport and specific heat measurements performed in mm-sized LuPdBi single crystals. Because of the low bulk mobility, it was revealed that our samples displayed a two-dimensional (2D) like weak antilocalization (WAL) effect, one of the hallmarks of topological surface states, as inferred from an angle-dependent magnetoconductance study and quantitative analysis with 2D Hikami-Larkin-Nagaoka (HLN) theory. This provides the first experimental evidence of the signature of the surface states in the transport properties in half-Heusler compounds, confirming that the LuPdBi half-Heusler compound forms a topological nontrivial semimetal. In addition, the LuPdBi displays bulk superconductivity with a critical temperature of $T_c \sim 1.7$ K, this being the highest $T_c$ reported so far in half-Heusler compounds. The nature of bulk superconductivity and the transport signature of topological surface states reported here suggest strongly that the LuPdBi system represents a promising candidate for realizing a topological superconductor associated with intriguing Majorana edge states.

## Results and discussion

Single crystals of LuPdBi were grown by a flux method as described in the Methods section. **Figure** 2a shows a representative powder X-ray diffraction pattern of the ground LuPdBi crystals. All peaks are well indexed to a MgAgAs-type structure with a lattice constant of 6.63 Å as depicted in the inset, indicating that the samples crystallize in a single phase without impurity. The Laue back-reflection pattern shown in Figure 2b indicates an apparently well developed (100) planes, further confirming the high quality of our single crystal. In addition, we found that all the signals collected by EDX are from three elements (Lu, Pd and Bi) and an elemental mapping (Figure 2b) obtained by using the EDX demonstrates the uniform distribution of the elements Lu, Pd and Bi across the whole sample.

Figure 3a shows the temperature dependence of longitudinal resistivity $\rho_{xx}$ of a LuPdBi single crystal obtained with standard four-terminal geometry from 300 to 0.3K in a zero magnetic field. The values of electrical resistivity fall within the range of 0.4 to 1 mΩ cm, similar to that in other members of RPdBi series,[31] and can be ascribed to the characteristics of a semimetal or narrow band semiconductor. Although the resistivity increases monotonically upon cooling from 300 to 2 K, the behavior of the temperature dependences varies in the different temperature ranges. This nonmetallic resistivity behavior was also found in the $Bi_2(Se,Te)_3$ topological system with large bulk resistivity.[23,24] When further decreasing the temperature, the resistivity suddenly drops to zero, this being indicative of a superconducting transition at 1.7 K (see lower inset of Figure 3a). A close analysis of the temperature dependence of the resistivity reveals the following: as shown in the upper inset of Figure 3a, in a temperature range from 300K down to 90K, the electric conductance can be understood as a parallel circuit of an insulating bulk and a metallic surface. The insulating behavior of the bulk can be described by the three-dimensional (3D) variable-range hopping model (VRH): $\sigma_{xx} \propto \exp[-(T/T_0)^{-1/4}]$ .[32] When the temperature is below 90K, the resistivity tends to saturate and to deviate dramatically from the insulating behavior described by the 3D VRH model, which indicates strongly that a metallic transport channel plays a dominant role at low temperatures.

The Hall resistivity $\rho_{xy}$ was measured in the temperature range from 300 to 2 K with a magnetic field up to 10 T. Shown in the inset of Figure 3b is representative data obtained at 2 and 300K, respectively. The Hall resistivity curve can bend at higher fields when there are two kinds of charge carriers whose numbers are comparable participate in conduction. Howver, in our case, the Hall resistivity is almost linear in the whole field range, implying that the bulk contribution is still dominated. The positive slope and linear dependence of $\rho_{xy}(H)$ on the magnetic field indicate that one type of charge carrier (hole) dominates the transport properties. The carrier density $n$ and Hall mobility $\mu_H$ are then deduced based on the one-carrier model and the temperature dependences are presented in Figure 3b. The deduced values of $n$ and $\mu_H$ fall within the range of 4.9~10.0 × $10^{19}$ /cm$^3$ and 300~400 cm$^2$/(V s), respectively, and decrease with decreasing temperatures. It should be noted that the deduced carrier mobility in LuPdBi is significantly lower than that in other half-Heusler TIs,[12,14,22] which facilitates the detection of the signature of surface states in the transport properties of the samples.

To understand the origin of the conductive channel, we measured the magnetoresistance (MR) of LuPdBi at different temperatures. Remarkably, a pronounced cusp of MR was observed in the low magnetic field region, as shown in the normalized MR profiles from 2K to 200K in Figure 4a. This observation is reminiscent of the weak antilocalization (WAL) effect that has widely been reported in Bi-Se-Te TIs.[24,25,33,34] As a quantum correction to classical conductivity, WAL can result from both the strong spin-orbit coupling in the bulk of sample and the topological surface states of TIs. To further clarify the origin of the WAL, MR measurements were performed by varying the direction in which the magnetic field was applied to the sample. The variation of magneto-conductivity, $\Delta\sigma = \sigma(B) - \sigma(B = 0)$, obtained at 10K as a function of the perpendicular component of the applied magnetic field, $B\sin\theta$, is shown in Figure 4b. It is evident that the low field data (-0.3T to 0.3T) overlaps onto a universal curve, suggesting that the magneto-conductivity depends mainly on the vertical component of the applied magnetic field. These data indicate strongly a 2D nature of the electrical transports that are likely to originate from the surface channel. If the WAL is caused mainly by spin-orbit coupling in a 3D bulk channel, the magneto-conductivity should be independent of the tilt angles of the magnetic field.

To gain a deeper understanding of WAL phenomena, a more quantitative analysis is necessary. The effect of localization on the 2D magneto-conductivity can be described by the Hikami-Larkin-Nagaoka (HLN) theory.[35] In the limit of strong spin-orbit coupling and low bulk mobility regime, the weak field conductance variation can be written as :

$$\Delta\sigma(B) = A[\psi(\frac{1}{2} + \frac{\hbar}{4eBL_\varphi^2}) - \ln(\frac{\hbar}{4eBL_\varphi^2})] , \qquad (1)$$

Here, $\psi(x)$ is the digamma function, $L_\varphi$ is the phase coherence length, and the parameter $A = \alpha e^2 / 2\pi^2\hbar$ , for which in the thin film case $\alpha$ were equal to -1/2 (per surface) for WAL. By applying Eq. (1) to the experimental data, the fitting parameters of $L_\varphi$ and $A$ are obtained. In Figure 4c, we show a representative fitting result of the magneto-conductivity curve obtained at 10K in the weak field range -0.3T to 0.3T. It is clear that the data can be fitted well using the 2D HLN model. The temperature dependences of the $L_\varphi$ and $A$ values are summarized in Figure 4d. It should be noted that the value of $A$ in the order of $10^0$ $\Omega^{-1}$, about $10^5$ times larger than the theoretical 2D values. This disparity is most likely caused by the contribution from the dominated 3D bulk channels, which was also observed in another macroscopic system of Be$_2$Se$_3$ crystals.[36] In addition, the nearly constant value of $A$ with varying temperature indicates fixed constitution of conduction channels. The existence of 2D WAL was further confirmed by examining the temperature dependence of coherence length $L_{\varphi}$ as shown in Figure 4d. According to the WAL of TIs,[37,38] both the electron-electron (*e-e*) scattering and electron-phonon (*e-ph*) scattering are supposed to emerge in the 3D TIs. Therefore, the $L_{\varphi}$ as a function of temperature may be expressed as：

$$\frac{1}{L_{\varphi}^{2}(T)} = \frac{1}{L_{\varphi}^{2}(0)} + A_{ee}T^{p'} + A_{ep}T^{p}, \qquad (2)$$

Here, $L_{\varphi}(0)$ represents the zero-temperature coherence length, $A_{ee}T^{p'}$ and $A_{ep}T^{p}$ represent the contribution from the *e-e* and *e-ph* interaction, respectively. We fitted Eq. (2) to the values of $L_{\varphi}(T)$ obtained from the fitting of Eq. (1) to the magneto-conductivity measured at different temperatures, as shown by the red solid line in Figure 4d. The best fit yields $A_{ee} = 2.64\times10^{-7}$ /nm$^{2}$ K, $A_{ep} = 3.05\times10^{-8}$ /nm$^{2}$ K$^{2}$. It is apparent that the temperature dependence of $L_{\varphi}(T)$ can be well described by Eq. (2) with $p' = 1$ and $p = 2$ and $L_{\varphi}(0) = 94$ nm, which indicates clearly that a 2D magneto-transport channel participates in the transport of our LuPdBi crystals.

Other strong evidence for the existence of a 2D magneto-transport channel in LuPdBi crystals is the observation of unusual conductance fluctuations under high magnetic fields.[39] As shown in Figure S2, although no fixed period can be identified in the fluctuation, the gradual increasing of the amplitude with decreasing temperature distinguishes it from a random noise signal. Recently, universal conductance fluctuations (UCF) have been reported in Bi-Se-Te TIs which are likely to originate from their surface state transport.[40-42] As discussed in the supplementary, the conduction fluctuations owned a 2D feature that relates to the surface state, but the fluctuation amplitude is still too large comparing to the UCF characterized value of $e^{2}/h$. Thus, we can conclude that the experimentally observed WAL in LuPdBi crystals is likely to result from a combination of both the WAL of the 2D surface channels and WAL of 3D bulk channels. Moreover, we have grown and measured two other half-Heusler LaPdBi and YPdBi single crystals. Neither any signature of WAL nor conductance fluctuations in LaPdBi [39] and YPdBi [22] were found in these two materials. This test experiment demonstrates that, in these gapless half-Heusler systems, reducing the mobility of the bulk carriers is an effective and simple way for probing, for example, the 2D surface transport properties in the presence of bulk conduction.

The low temperature resistance data as measured under different magnetic fields is shown in the inset of Figure 5a. It is clear that the crystal becomes a superconductor below 1.7K. To our knowledge, 1.7K is the highest critical temperature that has been observed in the family of half-Heusler alloys or in the noncentrosymmetric system. To gain a deeper understanding of the physics associated with superconductivity, we extracted the upper critical field $B_{c2}(T)$ from the resistance curves, as shown in Figure 5a. The value of the upper critical field at absolute zero temperature, $B_{c2}(0) = 2.2$T was obtained by fitting the generalized Ginzburg-Landau model, $B_{c2}(T) = B_{c2}(0)(1-t^2)/(1+t^2)$, to the extracted data, where $t = T/T_c$. Using the relation $B_{c2} = \Phi_0/2\pi\xi^2$, where $\Phi_0$ is the flux quantum, we obtained superconductor coherence length, $\xi = 12$ nm. The mean free path of the charge carrier $\ell$ was calculated based on $\ell = \hbar\kappa_F/\rho_0 ne^2$, by assuming a spherical Fermi surface with wave number $\kappa_F = (3\pi^2 n)^{1/3}$. Taking the experimental values $n = 4.9 \times 10^{25}$ m$^{-3}$ and $\rho_0 = 4.2 \times 10^{-6}$ Ωm (Fig. 3(b)), we obtained $\ell = 23$ nm. Thus, $\ell > \xi$, which confirms that LuPdBi is sufficiently pure to allow for odd-parity SC.[43] We also compared the experimental data with the universal curve for a clean spin-singlet SC with an orbital limited upper-critical field $B_{c2}^{orb}(0) = -0.72 \times T_c[dB_{c2}/dT]_{T_c}$ [Werthamer-Helfand-Hohenberg (WHH) approximation].[44] At low temperatures, the experimental values of $B_{c2}(T)$ are clearly larger than those predicted by the WHH model. Actually, the $B_{c2}(T)$ curve exhibits a nearly linear dependence over the whole temperature range, as for YPtBi [12] and ErPdBi [15]. Based on the linear dependence, the zero temperature upper critical field was obtained $B_{c2}(0) = 2.6$T by simple extrapolation. In both cases, the value of $B_{c2}(0)$ for LuPdBi is well below the Pauli paramagnetic limit of $B_{c2}^p(0) = 1.85T_c \approx 3.1$ T, suggesting an orbital pair-breaking mechanism for LuPdBi.

To further explore the superconductivity properties of LuPdBi, we measured the specific heat $c_p(T)$ as a function of temperature under zero and 3T magnetic field. The experimental data are presented in the inset of Figure 5b as $c_p(T)/T$ vs. $T$. A jump appearing at 1.7K reflects the superconducting transition. The 3T curve corresponds to the normal state, because the superconductivity is completely quenched under 3T magnetic field. The normal state data were fitted to the Debye model, $c_p = c_e + c_{ph} = \gamma_n T + A_3 T^3 + A_5 T^5$, and the fitting curve is also plotted in the same curve (the green line). The first term in the Debye model represents the electron specific heat, and the latter terms are due to phonon contribution. From the fitting, $\gamma_n = 11.9$ mJ/mol K$^2$, $A_3 = 1.0$ mJ/mol K$^4$, and $A_5 = 0.36$ mJ/mol K$^6$ are obtained. The Debye temperature $\Theta_D$ was estimated to be 295 K. Subtracting the phonon contribution from the zero-field data gives the electronic specific heat $c_e$ in the SC state, as plotted in Figure 5b. The sharp-jump in the electron specific heat strong indicates that the superconductivity in LuPdBi is fully gapped and bulk in nature.

As discussed by Kriener *et al.*[45] for the topological superconductor Cu$_x$Bi$_2$Se$_3$,[46] we analyzed the $c_e/T$ data in the SC state as shown in Figure 5b. We attempted to fit the standard BCS model to the $c_e/T$ data shown in Figure 5b and found that the data cannot be described by the standard BCS model (fitting not shown here). We then attempted to fit the modified BCS model proposed for strong-coupling superconductors[47] to the experimental data. The total specific heat $c_e$ contains two parts: $c_{es}$ (superconducting state) and $c_{en}$ (normal state), which can be expressed as follows:

$$c_{es}/T = \frac{4N(0)}{k_B T}\int_0^{\hbar w_D} d\varepsilon[\varepsilon^2 + \Delta^2 - \frac{T}{2}\frac{d\Delta^2}{dT}](1 + e^{E/k_B T})^{-2}e^{E/k_B T} + \gamma_{res}, \quad T \le T_c \quad ,$$

$$c_{en}/T = \gamma_n, \qquad\qquad\qquad\qquad\qquad\qquad\qquad\qquad\qquad T > T_c \qquad (3)$$

where $E = \sqrt{\varepsilon^2 + \Delta^2}$, $\Delta(T)$ is the temperature dependent gap function, $w_D$ is the Debye frequency and $\gamma_{res}$ represents a constant term account for contribution of the non-SC part. In the modified BCS model, $\alpha = \Delta(0)/k_B T_c$ is taken as an adjustable parameter. By fitting the above model to the experiment data $c_e/T$, as clearly seen in Figure 5b, it can be reasonably reproduced by the modified BCS model (red line) with $\Delta(0) = 0.32$ meV and $\alpha = 2.6$. The $\alpha$ value is larger than the weak-coupling BCS value of 1.76, indicating LuPdBi is a strong-coupling superconductor. We should point out that the pairing symmetry in LuPdBi may not be the simple isotropic $s$ wave as indicated by Kriener *et al.* [45] for the topological superconductor $Cu_xBi_2Se_3$. The fitted parameter for $\gamma_{res}$ is 9.5 mJ/mol K$^2$, thus the value of $\gamma_s(= \gamma_n - \gamma_{res})$ for SC is 2.4 mJ/mol K$^2$. It may be noted further that the carrier density $n$ obtained from Hall measurements is $4.9\times10^{19}$ /cm$^3$ while the value of $\gamma_n$ is 11.9 mJ/mol K$^2$ for LuPdBi single crystal. This result seems to be inconsistent with the results obtained in $Cu_xBi_2Se_3$, where $n = 1.3\times10^{20}$ /cm$^3$ and $\gamma_n = 1.95$ mJ/mol K$^2$. The main reason for this disparity is likely to be due to the difference in the effective mass of the charge carriers. In our LuPdBi single crystal, the effective mass of the carriers $m^*$ is estimated to be $9.5\,m_e$ ($m_e$ being the mass of free electron), which is much larger than that of the topological superconductor $Cu_xBi_2Se_3$.[46] Of course, further experimental and theoretical work will be necessary to reveal more details of the large mass enhancement in LuPdBi.

In summary, a thorough study has been undertaken of the magneto-transport and specific heat properties of half-Heusler LuPdBi single crystals. A clear 2D WAL effect and universal conductance fluctuations, both reflecting the existence of topological surface states, were observed experimentally in mm-sized LuPdBi crystals. This is, to our knowledge, the first experimental observation of the transport signature of surface states in the proposed half-Heusler TIs, which can be ascribed to the low bulk mobility. Although 2D WAL was observed, due to the large contribution of the 3D bulk states, the bulk contribution can not be excluded when connected this effect to the high mobility topological surface states. It should be pointed out that the conductance quantization was invisible due to the absence of the band gap opening in the bulk. The band gap in the bulk could be opened by doping or applying pressure.[6-8] Besides the observation of a 2D WAL effect, a superconducting transition was revealed at $T_c$ ~1.7 K in the bulk LuPdBi. This is probably the highest transition temperature in the family of half-Heusler alloys or in the noncentrosymmetric system. An analysis in the framework of a generalized BCS theory leads to the conclusion that the superconductivity of noncentrosymmetric LuPdBi is fully gapped with a possibly unconventional pairing character. It will be interesting to confirm the fully gapped and odd-parity SC using other techniques. These results suggest strongly that LuPdBi is a promising candidate as a topological superconductor for realizing Majorana fermions for future application in quantum computation.

## Methods

**Sample preparation.** Single crystals of LuPdBi were grown by using the self-flux method. Proper ratios of high purity metals of lutetium (99.95%), palladium (99.99%), and bismuth (99.999%) were mixed and sealed in Tantalum foil and then annealed in evacuated quartz tubes at 1073 K for two weeks. The second step was to grow crystals of LuPdBi by a flux method. The constituent elements were in a high-purity argon atmosphere. Lu (99.95%) pieces, Pd (99.99%) grains, and Bi (99.999%) grains were used as starting materials. The fabricated polycrystalline LuPdBi was grounded, mixed with fluxed Bi powders in an atomic ratio of 1:10, and placed in a Tantalum crucible, which was loaded into a fused quartz tube. For estimation of the best ratio of the fluxed material to LuPdBi, we examined the mixtures with various weight ratios, ranging from 20:1 to 1:1. The 10:1 mixture resulted in the best solubility and crystal growth. The tube was sealed under Ar gas at the pressure of $10^{-4}$ Torr and then placed in a furnace. We also tried to estimate the best cooling rate. The best growth conditions at the 10:1 ratio were as follows: The mixed sample was heated from room temperature to 1150 ℃ for 24 hours, and then keeped at 1150 ℃ for 24 hours and finally the sample was slowly cooled to 850 ℃ at a rate of 2 ℃∕hour.

**Materials characterization.** The composition of the single crystal samples was determined by energy-dispersive X-ray (EDX) spectroscopy and the structure were checked by X-ray diffraction (XRD) with Cu-$K\alpha$ radiation. The single-crystal orientation was checked by a standard Laue diffraction technique.

**Transport measurements.** To measure the transport properties of a single crystal (1.5×1.5×0.2 $mm^3$), the electrical leads in the four-probe and Hall bar configurations were attached onto the sample using silver paste by gold wires. The transport and specific heat measurements were performed in the temperature range from 0.4 to 300 K on a Quantum Design PPMS-9 system with a $^3$He refrigeration insert. Electrical leads were attached to the samples using room-temperature cured silver paste by gold wires. The perpendicular and longitudinal magnetoresistances were measured with the magnetic field perpendicular or parallel to the electrical current direction. Magnetoresistance versus magnetic field was also measured for several temperatures. The Hall effect was measured by rotating the crystal by 180° in a magnetic field of 10 T and the Hall coefficient was calculated from the slope of the measured Hall effect curves.

**Band structure calculations.** The electronic structure calculations in this work were performed using full-potential linearized augmented plane-wave method, as implemented in the package WIEN2K. The exchange correlation of electrons was treated within the local spin density approximation (LSDA) including spin orbit coupling (SOC). Meanwhile, a 17×17×17 $k$-point grid was used in the calculations, equivalent to 5000 k points in the first Brillouin zone. Moreover, the muffin-tin radii of the atoms are 2.5 a.u, which are generated by the system automatically. The lattice parameters and atomic positions were taken from our experimental data.

## Acknowledgements

We thank Prof. L. Shan at Institute of Physics, CAS, Beijing, for helpful discussions, and are acknowledge the continuous support provided by Dr. Y.G. Shi at Institute of Physics, CAS, Beijing, for helping with single crystal growth. This work was supported by funding from the "973" Project (2012CB619405) and NSFC (Nos. 51171207, 51025103 and 11274371).

## Author contributions

W. H. W and Z. Y. Liu designed the research project. G.Z.X., Y.D. and W.H.W. performed all the experimental measurements. E.K.L. and X.M.Z. performed band structure calculations. S.G.W., G.H.W. and X.X.Z. helped with the results analysis. G.Z.X. and W.H.W. wrote the manuscript. All authors reviewed the manuscript.

## Additional information

The authors declare no competing financial interests.

**Figure captions:**

**Figure 1 | Basic electronic structure of LuPdBi.** **(a)** The electronic band structure of LuPdBi with spin-orbital coupling. $\Gamma_8$ and $\Gamma_6$ band is highlighted by blue and red line. The heighted region by ellipsoid around $E_F$ corresponds to a Van Hove singularity along $\Gamma$–X direction; **(b)** Total and partial density of states with specific electrons. **(c)** A schematic illustration of the band inversion and hybridization.

**Figure 2 | Structural characterizations of LuPdBi single crystal.** **(a)** Powder XRD pattern of pulverized single-crystal LuPdBi sample (blue line), with the red line showing its calculation curve and the bottom trace displaying the intensity difference between the experimental and calculation data. Inset: crystallographic structure of LuPdBi. **(b)** Elemental maps of Lu (red), Pd (green), Bi (yellow) in a single crystal sample scanned by EDX. The typical mm-sized sample used for scanning, together with the Laue image, is shown in the left panel.

**Figure 3 | Transport properties of LuPdBi single crystal.** **(a)** Temperature-dependence of longitudinal resistivity $\rho_{xx}$ from 0.5 to 300 K. Lower inset highlights the superconducting transition occurring at $T_c$ =1.7 K in the enlarged view ranging from 0.5 to 5 K. Upper inset shows the longitudinal conductivity $\sigma$ in function of $T^{-1/4}$. Red line is the fitting of the 3D variable range hopping (VRH) behavior $\sigma \propto \exp[-(T/T_0)^{-1/4}]$ to the data, in the range of 90-300 K; deviation from the fitting at low temperature, marked by the arrow, signifies the parallel metallic conduction. **(b)** Temperature dependence of the carrier density $n$ (blue) and Hall mobility $\mu$ (red). Inset shows the linear behavior of the Hall resistivity $\rho_{xy}$ vs. $B$ at $T$ = 2 and 300 K, respectively.

**Figure 4 | Weak anit-localization (WAL) effect in LuPdBi single crystal.** **(a)** The normalized magnetoresistance ($R_{xx}$) in perpendicular magnetic fields ($B$) at temperature ranging from 2 to 200 K. It shows the WAL feature (sharp dips at around zero magnetic field) persistent until 100 K. **(b)** The normalized magneto-conductivity ($\Delta\sigma_{xx}$) plotted in the perpendicular magnetic field component of the magnetic field, $B\sin\theta$, at 10 K. The only perpendicular component dependent features indicate that WAL is induced by 2D surface electrons. Inset: schematic of the measurement setup where $\theta$ indicates the angle between the direction of the magnetic field and the current flow in the sample surface. **(c)** The change of magneto-conductivity ($\Delta\sigma_{xx}$) at 10 K (blue circle) and its fit to the 2D WAL mode (red line). **(d)** Temperature dependence of phase coherence length $L_\varphi$ deduced from the WAL fit under a field limit of 0.3 T (square). The red line shows the fit according to the Eq. (2) with $p' = 1$ and $p = 2$ . <span style="color:red">Inset shows the temperature dependence of the parameter $A$ in the Eq. (1).</span>

**Figure 5 | Superconducting properties of LuPdBi single crystal.** **(a)** The upper critical field $B_{c2}$, determined as the vanishing point of the superconducting state (inset), was plotted with the reduced transition temperature t=$T/T$c. The blue solid line shows the Ginzburg-Landau (G-L) fit to the data and the dotted green line represents the WHH theoretical curve. Inset: temperature dependence of resistance under different applied magnetic fields). **(b)** Temperature dependent electronic specific heat $c_e$ under zero field after subtracting the phonon contribution based on the Debye fit of the normal state (indicated by the inset). The red solid line represents the calculated $c_e/T$ curve given by strong-coupling BCS theory with $\alpha = \Delta(0)/k_B T_c = 2.6$, $\Delta(0)$ =0.32 meV, $T_c$=1.45 K. Inset: $c_p/T$ versus $T$ under 0 and 3T magnetic fields. The 3T magnetic field is larger enough to quench superconductivity, therefore the 3T curve represent the normal state of the material.

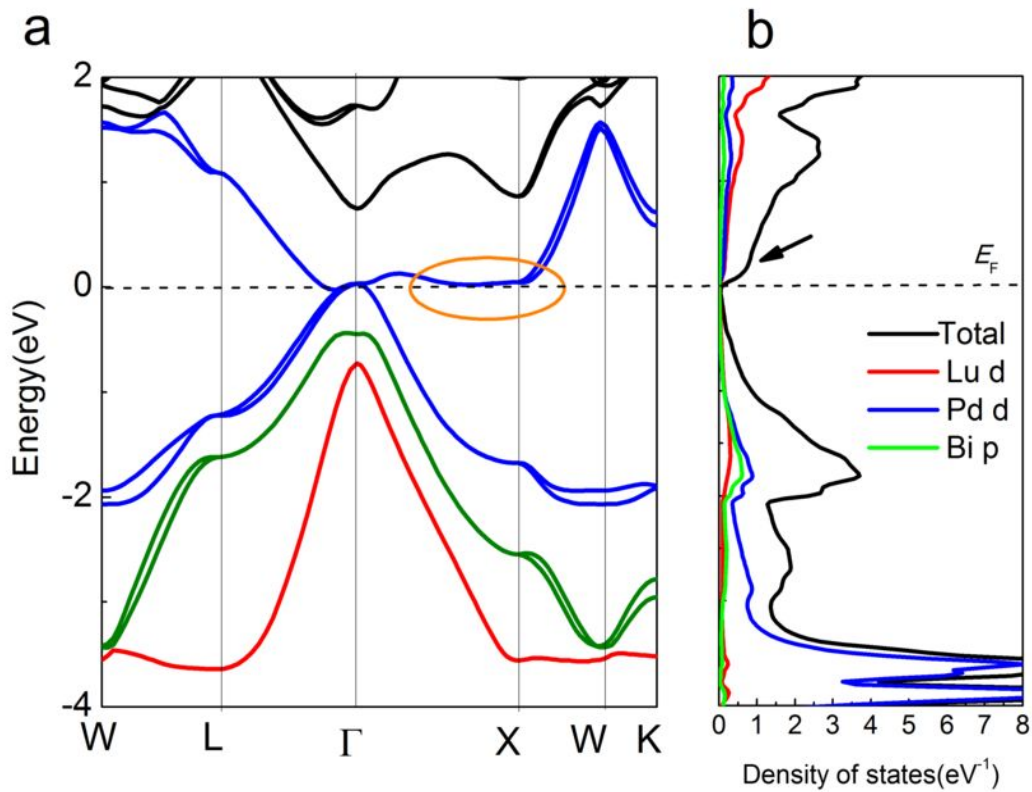

a

b

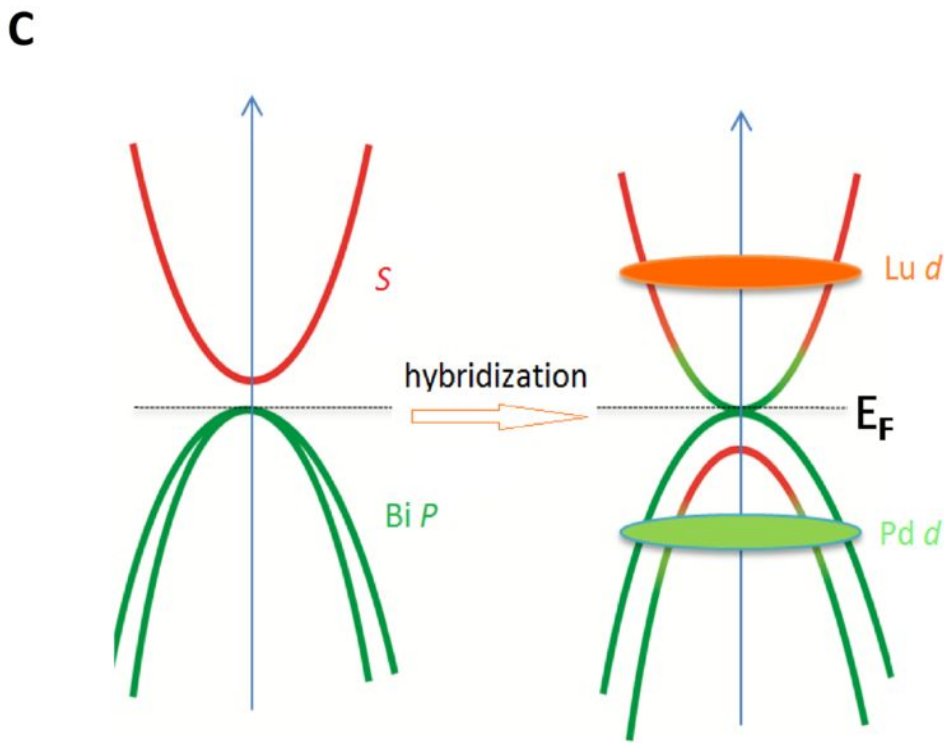

c

**a**

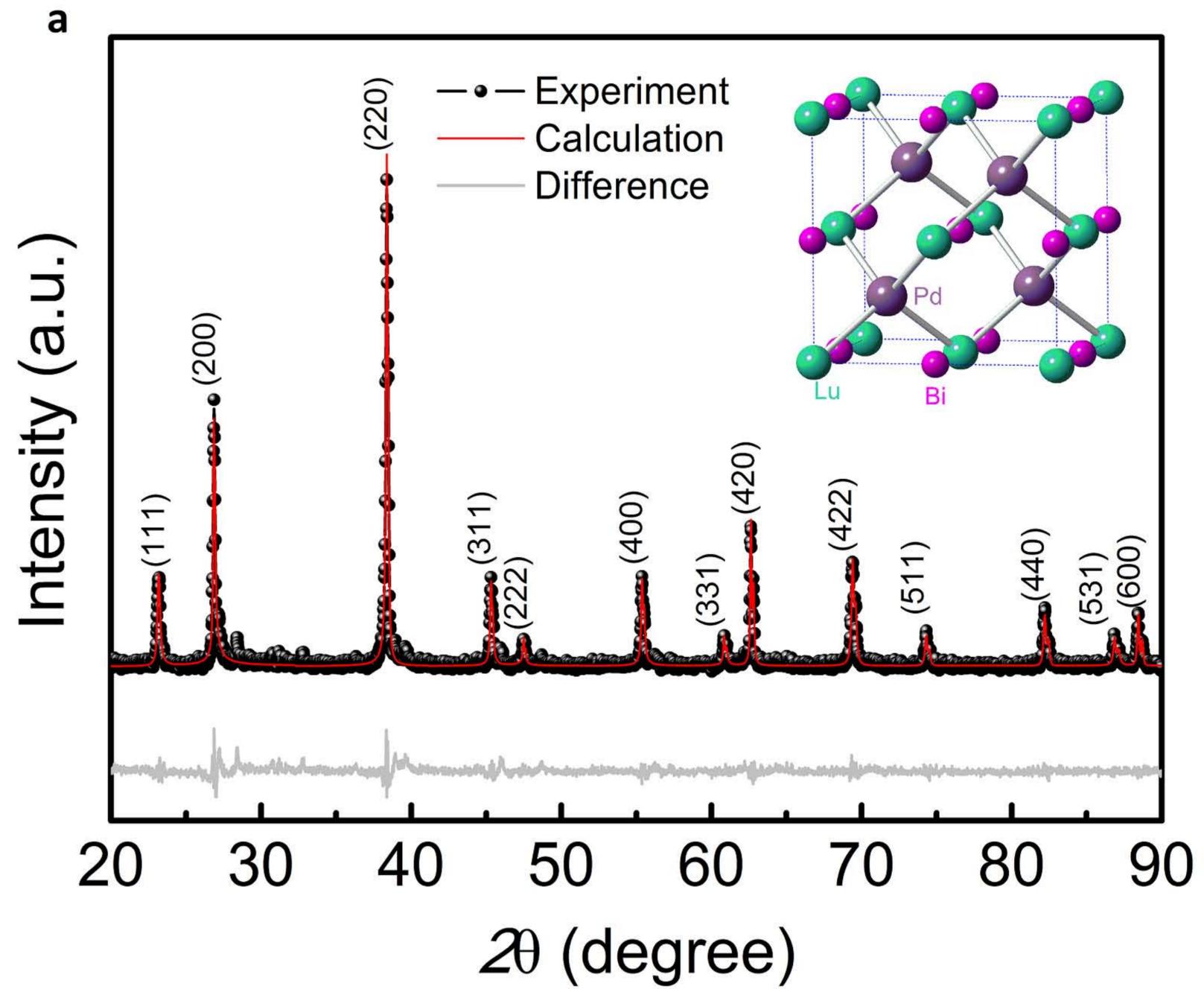

**b**

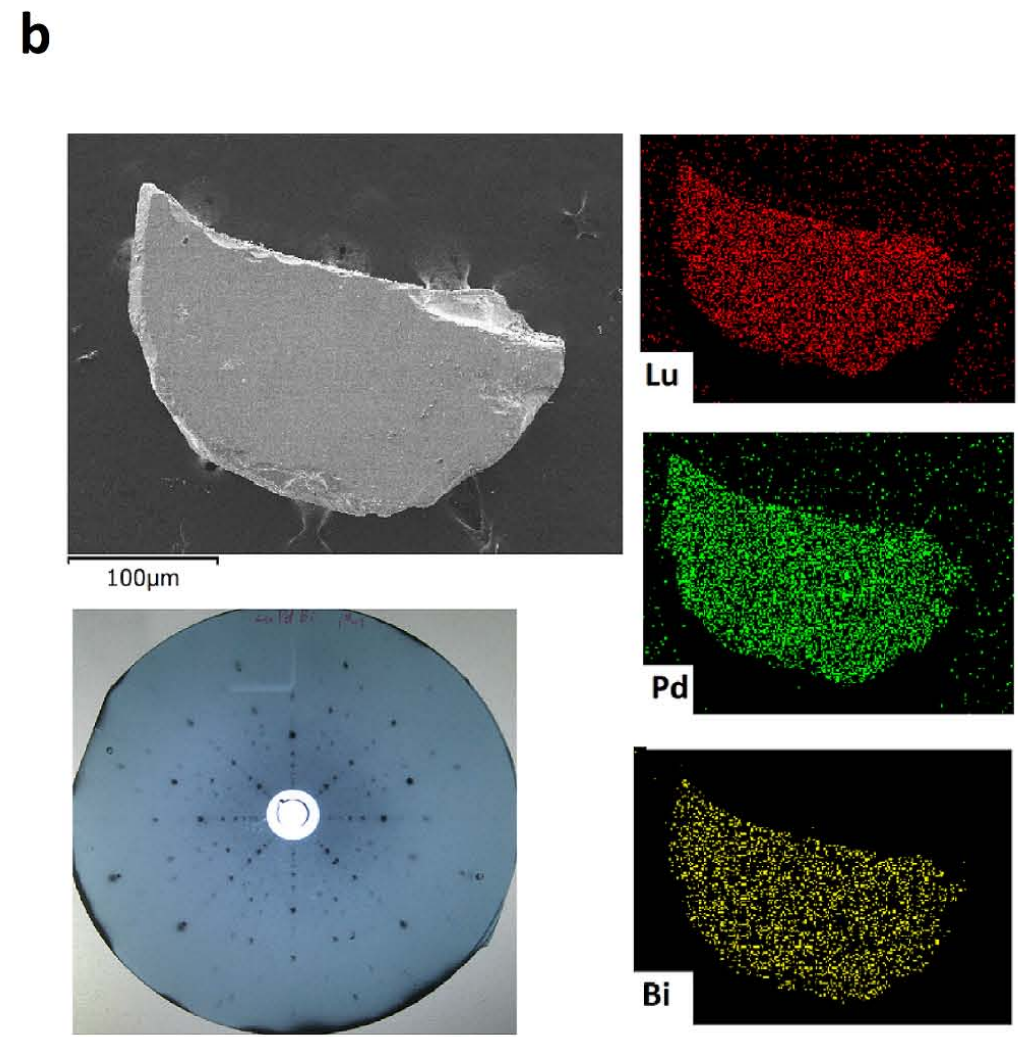

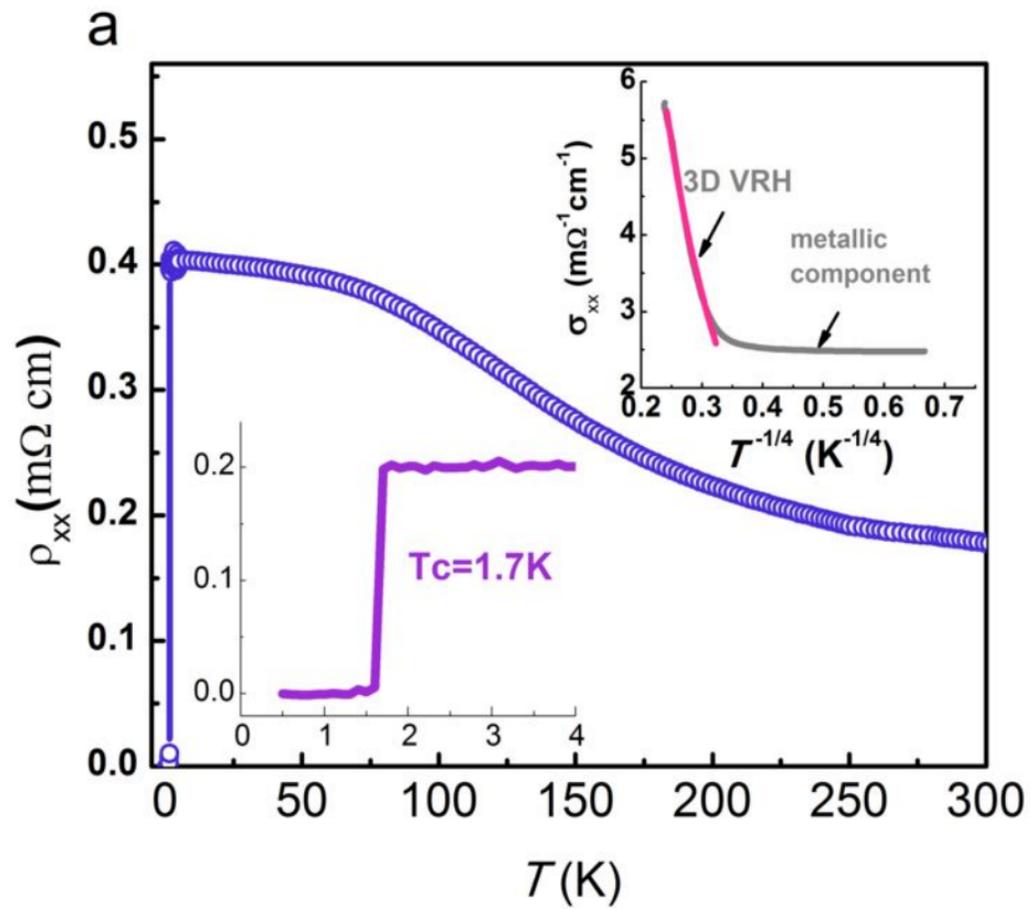

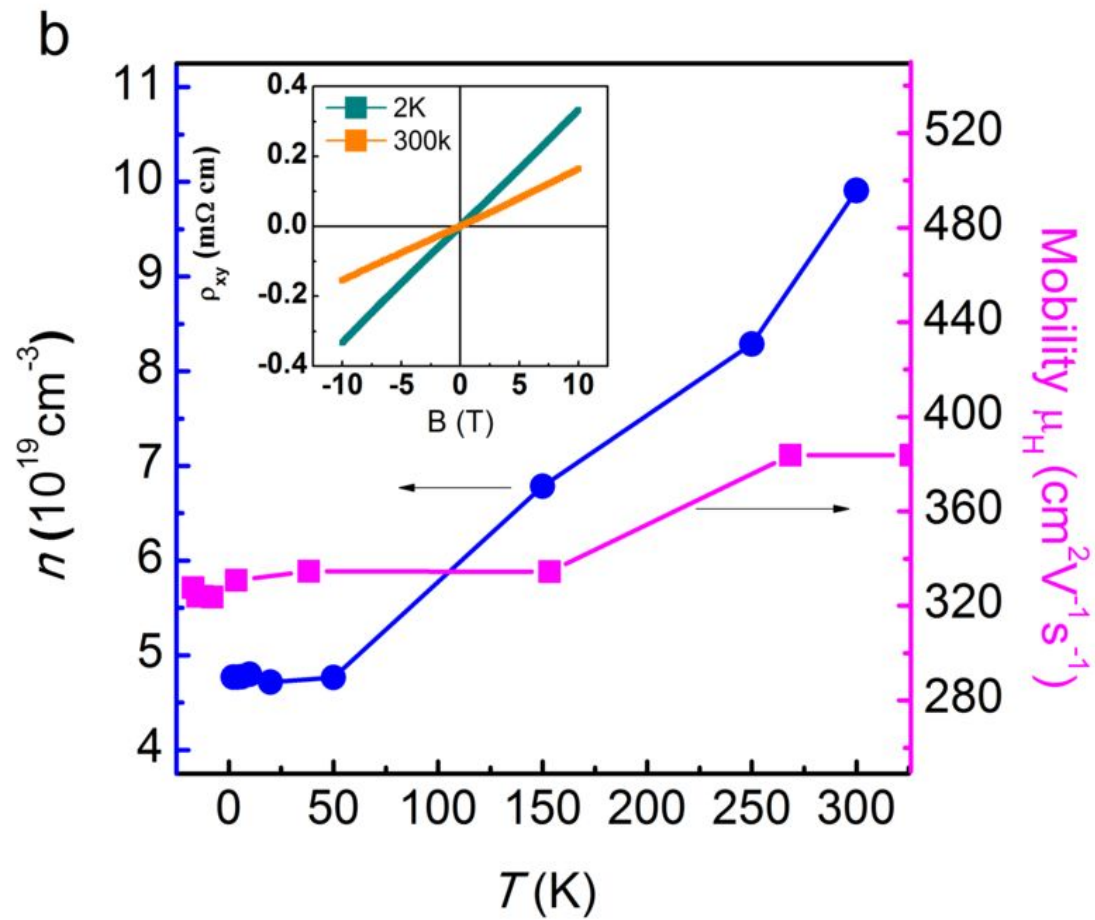

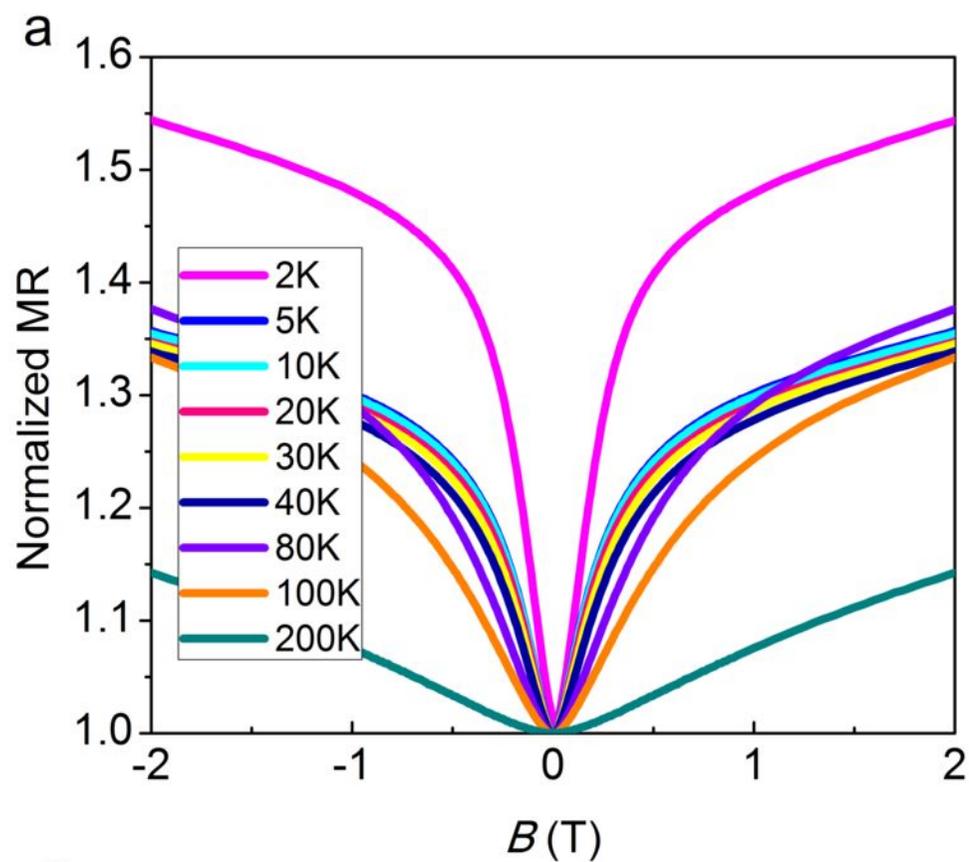

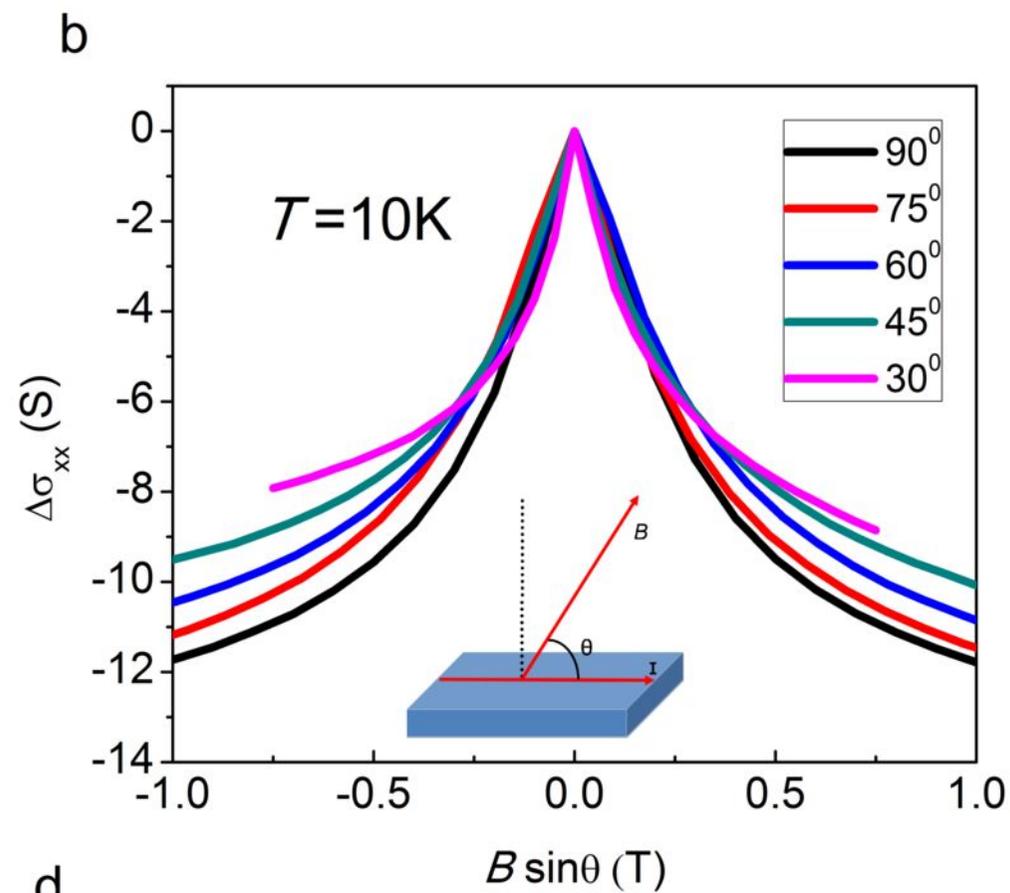

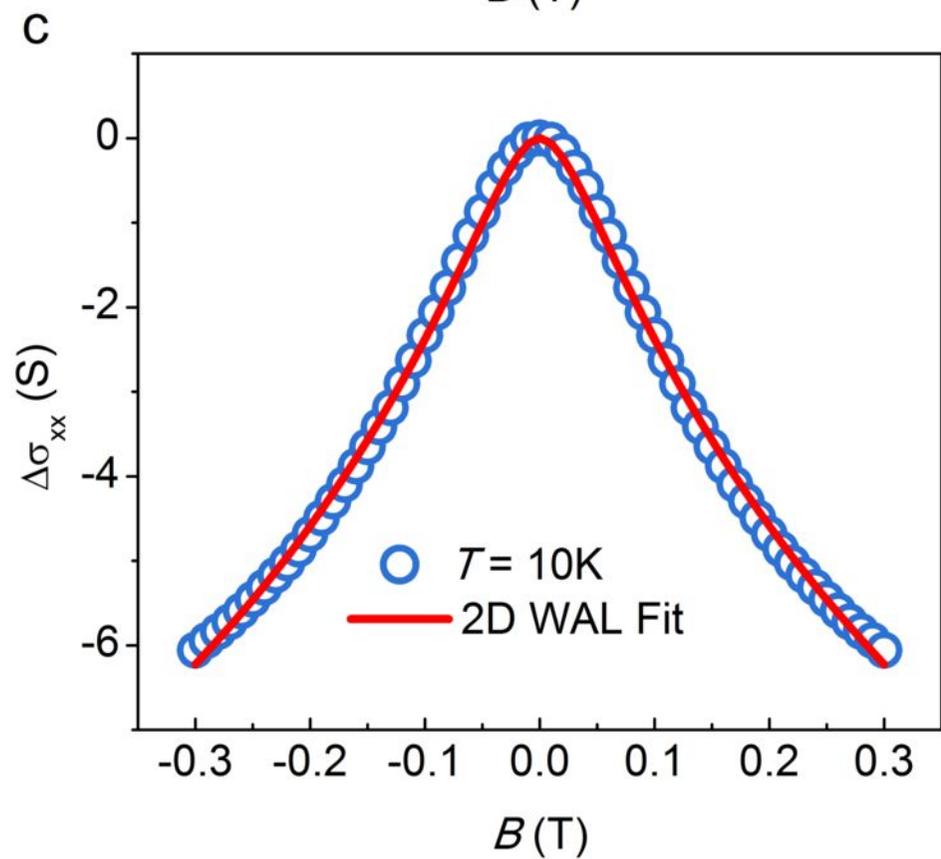

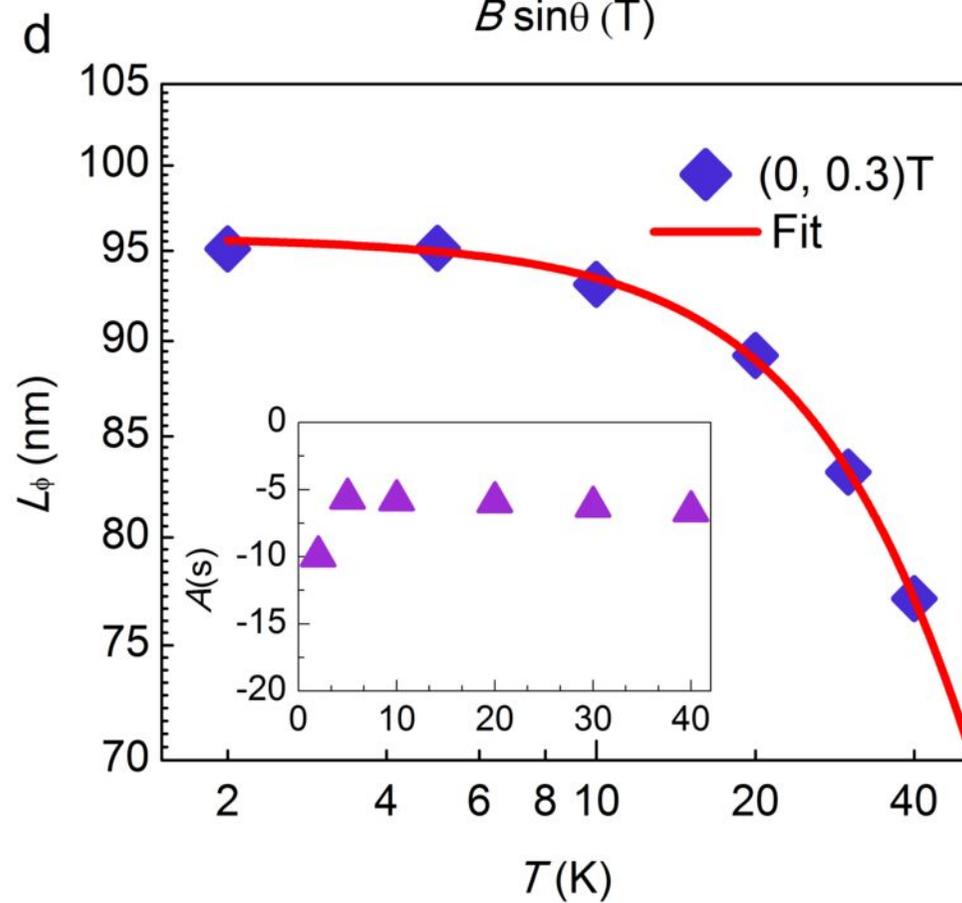

a

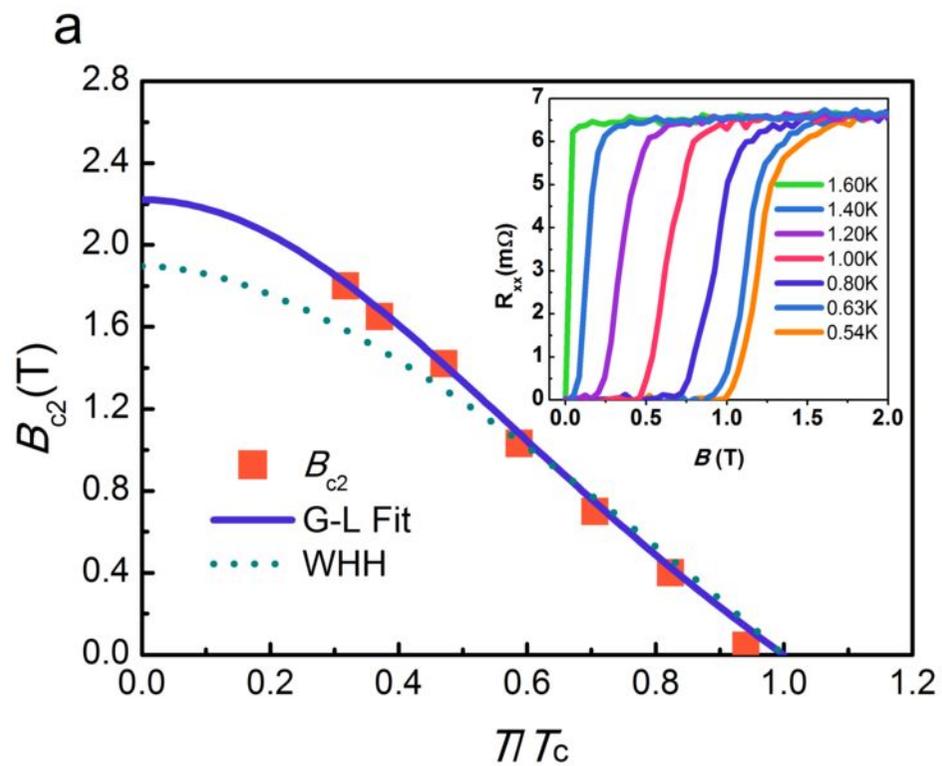

b

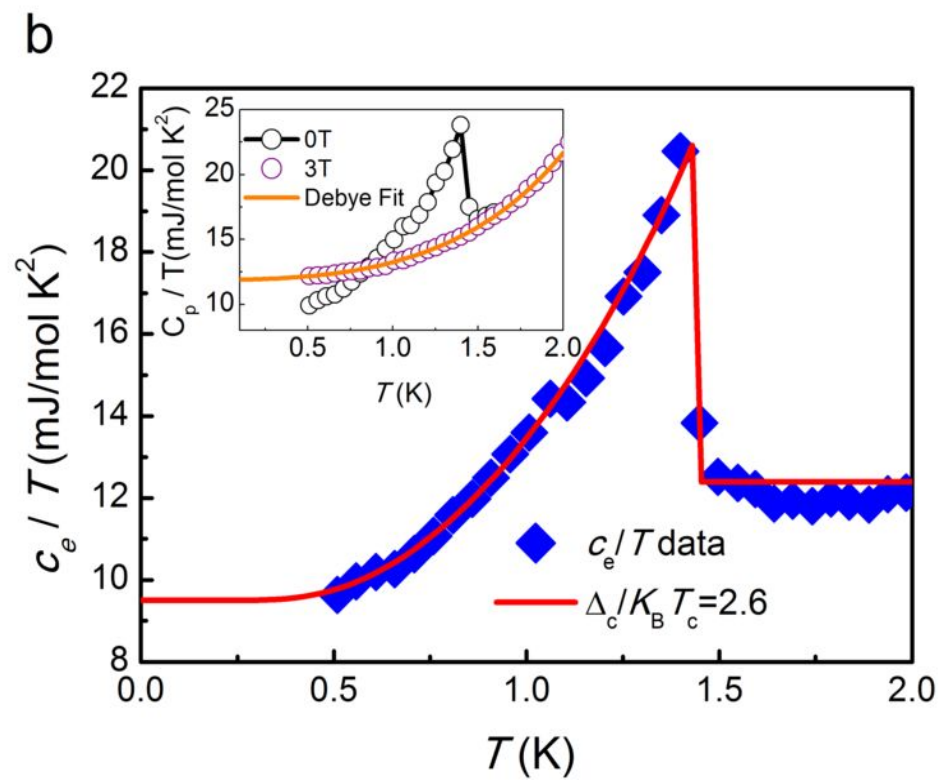